\documentclass[reprint,aps,prl,superscriptaddress,nobibnotes]{revtex4-1}%
\usepackage{amsmath}
\usepackage{amssymb}
\usepackage{natbib}
\usepackage{graphicx}
\usepackage{xcolor}
\usepackage{amsfonts}%
\providecommand{\U}[1]{\protect\rule{.1in}{.1in}}

\setcitestyle{numbers,square}
\begin{document}

\title{Non-Contact Spin Pumping by Microwave Evanescent Fields}
\author{Tao Yu}
\affiliation{Kavli Institute of NanoScience, Delft University of Technology, 2628 CJ Delft, The Netherlands}
\author{Gerrit E. W. Bauer}
\affiliation{Institute for Materials Research \& WPI-AIMR \& CSRN, Tohoku
	University, Sendai 980-8577, Japan} \affiliation{Kavli Institute of
	NanoScience, Delft University of Technology, 2628 CJ Delft, The Netherlands}
\date{\today}

\begin{abstract}
The angular momentum of evanescent light fields has been studied in
nano-optics and plasmonics, but not in the microwave regime. Here we predict
non-contact pumping of electron spin currents in conductors by the evanescent
stray fields of excited magnetic nanostructures. The coherent transfer of the
photon to the electron spin is proportional to the $g$-factor, which is large
in narrow-gap semiconductors and surface states of topological insulators. The spin pumping current is chiral when the spin susceptibility
displays singularities that indicate collective states. However, 1D systems
with linear dispersion at the Fermi energy such as metallic carbon nanotubes
are an exception since spin pumping is chiral even without interactions. 

\end{abstract}
\maketitle

\textit{Introduction}.---Efficient transfer of spin information among
different entities is a key objective in spintronics \cite{spintronics}. The
electromagnetic field at frequency $\omega$ carries a spin angular momentum
density \cite{Nori_Science,Nori_review}
\begin{equation}
{\pmb {\mathcal{D}}}=\frac{1}{4\omega}\operatorname{Im}\left(  \varepsilon
_{0}\mathbf{E}^{\ast}\times\mathbf{E}+\mu_{0}\mathbf{H}^{\ast}\times
\mathbf{H}\right)  ,
\end{equation}
where $\mu_{0}/\varepsilon_{0}$ are the vacuum permeability/permittivity, and
in the microwave regime the magnetic field component $\sim\mathrm{Im}%
(\mathbf{H}^{\ast}\times\mathbf{H})$ dominates the contribution of the
electric field $\mathbf{E}$. The evanescent fields at boundaries can have
local angular momentum even when the (linearly-polarized) propagating ones
have not \cite{Jackson,nano_optics}. A distinguishing feature of such
evanescent fields is the locking between the linear and angular momentum
\cite{Nori_Science,Nori_review,near_field,Jacob}. The chiral electrical
near-field of a rotating electrical dipole \cite{Jackson,nano_optics}, e.g.,
unidirectionally excites surface plasmon polaritons
\cite{near_field,Petersen,nano_optics}. Metallic striplines or coplanar
waveguides biased by currents in the GHz regime also emit chiral magnetic
near-fields \cite{poineering_1,poineering_2,Springer_book}, which is of
considerable interest for magnonics
\cite{magnonics1,magnonics2,magnonics3,magnonics4}, since chiral excitation is
a robust and switchable mechanism to pump a DC unidirectional magnon current
by an AC field \cite{pumping,nanowire}.

\begin{figure}[th]
\begin{center}
{\includegraphics[width=8.5cm]{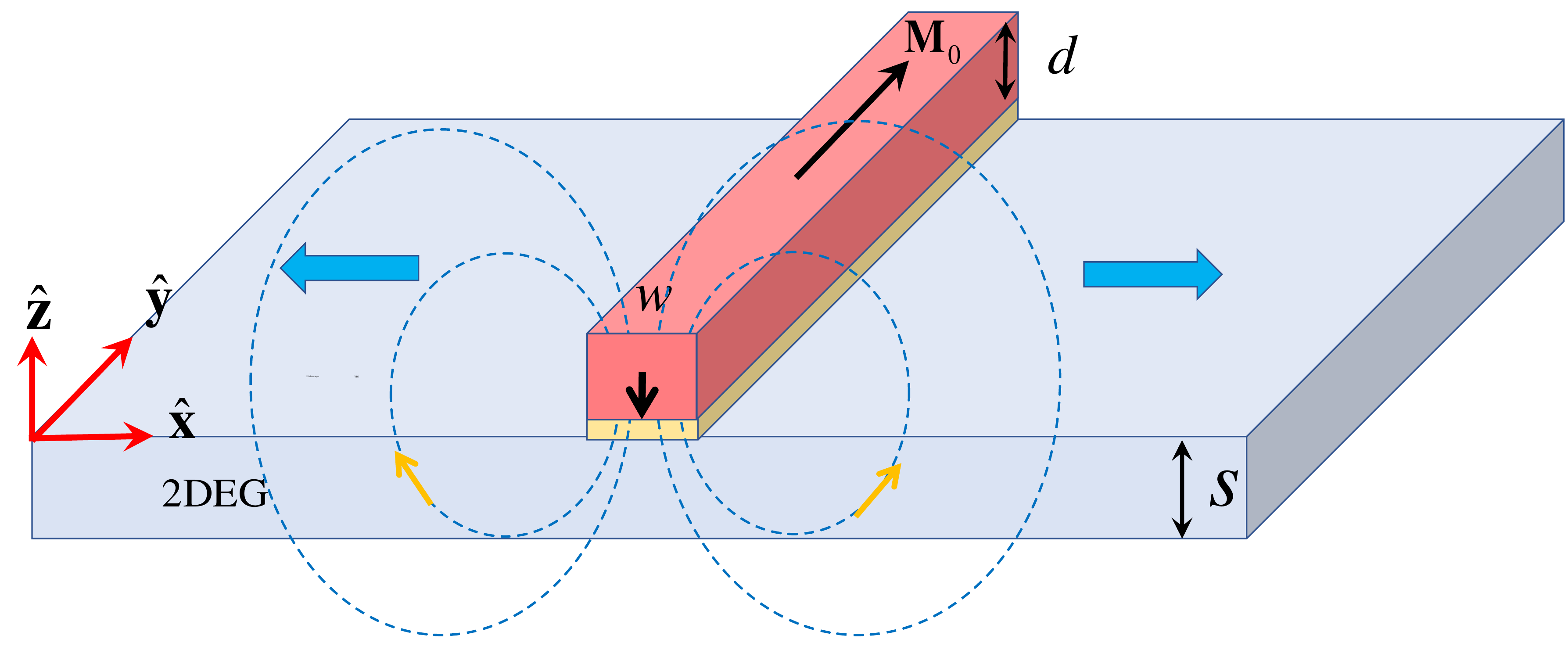}}
\end{center}
\caption{Snapshot of spin pumping by the microwave dipolar
field of an excited magnetic nanowire on top of a 2DEG. A thin tunneling
barrier suppresses any exchange coupling. Orange arrows indicate the direction
of stray field.}%
\label{model}%
\end{figure}

Spin pumping by exchange interaction is established when the magnet and
conductor form a good electric contact, which is difficult to achieve between
metals and semiconductors including graphene because of Schottky barriers and
electronic structure mismatch \cite{spin_pumping1,spin_pumping2}. Even when a
good contact to a magnet can be established, results may be difficult to
interpret due to proximity effects. Spin pumping at a distance by microwaves
solves these issues since it does not require direct contact between the
magnet and the system of interest. In this Letter, we address the non-contact
angular momentum transfer to an electric conductor by stray magnetic fields
emitted by an excited magnet, thereby generalizing the concept of spin pumping
by a contact exchange interaction \cite{spin_pumping1,spin_pumping2}. We are
motivated by the significant near fields that couple magnetic nanowires and
ultrathin magnetic insulating films, causing several chiral magnon transport
phenomena
\cite{transducer_simulation,Yu1,Yu2,Jilei_PRL,nanowire,accumulation1,accumulation2,spin_wave_diode}%
. Here we demonstrate that a magnetodipolar field pumps electron spins into a
conductor without need of electric contacts. We
illustrate the physics for a simple yet realistic model system of a magnetic
nanowire on top of a two-dimensional electron gas (2DEG) as illustrated in
Fig.~\ref{model}. The latter may be graphene
\cite{graphene_spin_pumping,YIG_graphene_exp,graphene_exp1,graphene_exp2}, but
the effect is strongly enhanced by spin-orbit interaction, such as a large
$g$-factor in InAs or InSb quantum wells (QWs)
\cite{g_factor_InAs,g_factor_InSb} or the surface states of 3D topological
insulator \cite{g_factor_Bi2Se3,YIG_topology}. In contrast to the dipolar spin
pumping of magnons, the spin pumping current in non-interacting conductors is
in general not chiral. However, the singular spin susceptibility in
one-dimensional systems with linear dispersion at the Fermi energy such as
metallic carbon nanotubes generates chirality of the spin injection with and
without interactions.

\textit{Transverse spin density of microwaves.}---We first demonstrate that
the evanescent magnetodipolar field of a magnetic nanowire carries transverse
angular momentum or \textquotedblleft spin\textquotedblright. The nanowire
with width $w$, thickness $d$ and equilibrium magnetization $\mathbf{M}_{s}$
along the wire $y$-direction, on top of an electron gas confined in the
$z$-direction on a length scale $s$ as illustrated in Fig.~\ref{model}, acts
as an antenna for external microwaves with frequency tuned to the ferromagnetic
resonance (FMR) $\omega_{\mathrm{K}}$ \cite{transducer_simulation,transducer_Dirk,transducer_Haiming}. In the following we use a quantum mechanical notation for convenience, but in
the classical limit operators can be simply replaced by field amplitudes. The
quantum formalism may form a starting point to study quantum effects in the
electron gas, such as spin-pumping-induced magnetic quantum noise
\cite{quantum_noise} at low temperatures or quantum squeezing and entanglement
of the electrons with microwaves via the magnet \cite{entanglement_Zou}. 

The magnetization dynamics expressed by the spin operator $\hat{\mathbf{S}%
}(\mathbf{r},t)$ generates a magnetic field by Coulomb's Law \cite{Landau},
\begin{equation}
\mathbf{H}_{\beta}(\mathbf{r},t)=-\frac{\gamma\hbar}{4\pi}\partial_{\beta
}\partial_{\alpha}\int d\mathbf{r}^{\prime}\frac{\langle\hat{\mathbf{S}%
}_{\alpha}(\mathbf{r}^{\prime},t)\rangle}{|\mathbf{r}-\mathbf{r}^{\prime}%
|},\label{dipolar}%
\end{equation}
in the summation convention over repeated spatial (or spin) indices
$\{\alpha,\beta\}=\{x,y,z\}$. $-\gamma$ is the gyromagnetic ratio of the
nanowire. For sufficiently weak excitation the spin operators in the wire can
be expanded into magnon field operators $\hat{\alpha}_{k_{y}}$ and their
amplitudes across the nanowire $m_{x,z}^{k_{y}}(x,z)$:
\begin{equation}
\hat{\mathbf{S}}_{x,z}(\mathbf{r})=\sqrt{2S}\sum_{k_{y}}\left(  m_{x,z}%
^{k_{y}}(x,z)e^{ik_{y}y}\hat{\alpha}_{k_{y}}+\mathrm{H.c.}\right)
,\label{magnon_operator}%
\end{equation}
where $S=M_{s}/(\gamma\hbar)$. The static stray field is negligibly small for
sufficiently long nanowires. The dynamic stray field $\mathbf{H}$ is the
response to $\langle\hat{\alpha}_{k_{y}}\rangle$, the coherent amplitude of
magnons with momentum $k_{y}\hat{\mathbf{y}}$ excited by external microwaves.
With Fourier components $\mathbf{H}_{\beta}(z,\mathbf{k},t)=\tilde{\mathbf{H}%
}_{\beta}(z,\mathbf{k})e^{-i\omega_{\mathrm{K}}t}+\tilde{\mathbf{H}}_{\beta
}^{\ast}(z,-\mathbf{k})e^{i\omega_{\mathrm{K}}t}$ for $\mathbf{k}=(k_{x}%
,k_{y},0)^{T}$, below the nanowire ($z<0$) \cite{Yu1,Yu2,nanowire},
\begin{align}
\left(
\begin{array}
[c]{c}%
\tilde{\mathbf{H}}_{x}(z,\mathbf{k})\\
\tilde{\mathbf{H}}_{y}(z,\mathbf{k})\\
\tilde{\mathbf{H}}_{z}(z,\mathbf{k})
\end{array}
\right)   &  =F_{\mathbf{k}}\left(  m_{z}^{k_{y}}+\frac{ik_{x}}{k}m_{x}%
^{k_{y}}\right)  \left(
\begin{array}
[c]{c}%
ik_{x}/k\\
ik_{y}/k\\
1
\end{array}
\right)  \nonumber\\
&  \times e^{kz}\left\vert \left\langle \hat{\alpha}_{k_{y}}\right\rangle
\right\vert ,\label{dipolar_field}%
\end{align}
where $F_{\mathbf{k}}=-\gamma\hbar\sqrt{2S}(1-e^{-kd})\sin(k_{x}w/2)/k_{x}$ is
the form factor of the rectangular wire.

\begin{figure}[th]
{\includegraphics[width=5.2cm]{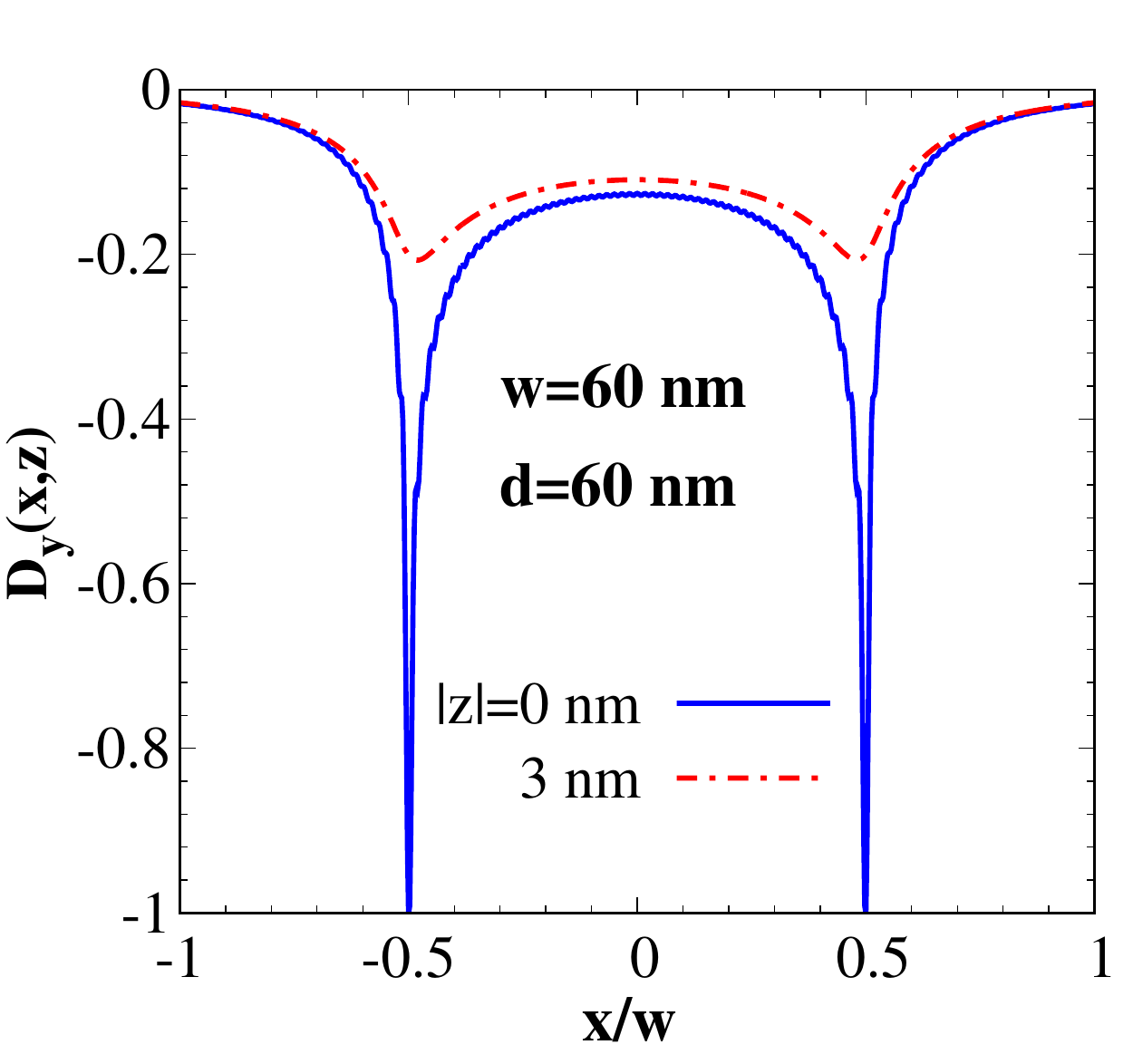}}\caption{Spatial distribution of the microwave photon spin density along the
$y$-direction (normalized by the maximum modulus when $z=0$) generated by a
magnetic nanowire with $d=w=60$~nm under FMR. The distances to the nanowire
are $z=0$ (solid line) and 3~nm (dashed-dotted line).}%
\label{transverse_spin}%
\end{figure}

$\tilde{\mathbf{H}}(z,\mathbf{k})$ decays exponentially $\sim e^{-k|z|}$ on a
scale governed by complex momentum $k_{x}\hat{\mathbf{x}}+k_{y}\hat
{\mathbf{y}}-ik\hat{\mathbf{z}}$ \cite{Jackson}. External microwaves excite
the Kittel magnon \cite{Kittel} with $k_{y}=0$, such that $\tilde{H}%
_{y}(\mathbf{k})$ vanishes and $\tilde{H}_{x}(\mathbf{k})=i\mathrm{sgn}%
(k_{x})\tilde{H}_{z}(\mathbf{k})$. The polarization of the Kittel mode is
governed by the shape anisotropy and applied magnetic field
\cite{footnote}. We focus here on circularly polarized spin waves with
$m_{x}\rightarrow im_{z}$ in nanowires with circular/square cross sections or
sufficiently large magnetic field. In this case, $\tilde{\mathbf{H}%
}(\mathbf{k})=0$ for $k_{x}>0$, since magnons precess preferentially in one
direction, which distinguishes the chirality found here from the
polarization-momentum locking in optics \cite{Nori_Science,Nori_review},
noting that chirality would vanish for linearly polarized spin waves. The
photon spin density under the nanowire ${\pmb{\mathcal{D}}}(x,z)=\mu
_{0}\mathrm{\mathrm{\operatorname{Im}}}\left[  \tilde{\mathbf{H}}^{\ast
}(x,z)\times\tilde{\mathbf{H}}(x,z)\right]  /(4\omega)$ is purely transverse
since ${\pmb {\cal D}}\cdot\mathbf{H}=0$  for arbitrarily polarized spin
waves. As illustrated in Fig.~\ref{transverse_spin} for $w=d=60$~nm,
${\pmb{\mathcal{D}}}$ is symmetric with respect to the center of the nanowire.
At finite distances from the wire the near-singularity at the edges is smeared
out, but the average amplitude remains significant. The photon magnetic field
couples to the electron spins by the Zeeman interaction. Absorption transfers
the photon spin over distances limited by the evanescent decay length in
contrast to conventional spin pumping, which happens directly at the interface.

\textit{Formalism}.---The photon field derived above can excite spins into any
conductor in its proximity. Here we illustrate the concept by a 2DEG with Hamiltonian $\hat{H}_{0}$ in which only the lowest subband with
envelope wave function $\phi(z)$ is occupied (see Supplemental Material
Sec.~III for a 1D quantum wire \cite{supplemental}). The Zeeman coupling
between the conduction electron spin $\hat{\mathbf{s}}$ and the evanescent
(near) field amplitude $\mathbf{H}(\mathbf{r},t)$ reads \cite{Landau,Jackson}
\begin{equation}
\hat{H}_{\mathrm{Z}}=\mu_{0}\gamma_{e}\int d\mathbf{r}|\phi(z)|^{2}%
\hat{\mathbf{s}}(\boldsymbol{\rho},t)\cdot\mathbf{H}(\mathbf{r},t),
\end{equation}
where $\mathbf{r}=\boldsymbol{\rho}+z\hat{\mathbf{z}}$ and $\gamma_{e}%
=-g_{e}\mu_{B}/\hbar$ with $\mu_{B}$ and $g_{e}$ being the Bohr magneton and
(effective) electron $g$-factor, respectively. In the strictly 2DEG limit
$|\phi(z)|^{2}\rightarrow\delta\left(  z\right)  $ and $\mathbf{H}%
(\boldsymbol{\rho},z)\approx\mathbf{H}(\boldsymbol{\rho},z=0)$. The excited
spin density in the linear response reads
\begin{equation}
s_{l,\alpha}(\boldsymbol{\rho},t)=-\mu_{0}\gamma_{e}\sum_{\omega,\mathbf{k}%
}e^{i\mathbf{k}\cdot\boldsymbol{\rho}-i\omega t}\chi_{\alpha\beta}%
(\mathbf{k},\omega)H_{\beta}(\mathbf{k},\omega),\label{linear_response}%
\end{equation}
where $\chi_{\alpha\beta}(\mathbf{k},\omega)$ are the elements of the spin
susceptibility tensor \cite{Mahan,Vignale,supplemental}.
$\mathbf{s}_{l}$ decays with the dipolar field on the scale of the wire width.
Assuming that the spin diffusion length, which can be of the order of
micrometers in 2DEGs \cite{Marie} (and even longer in graphene
\cite{graphene_diffusion}), exceeds the field decay length (tens of
nanometer), we can compute $\chi_{\alpha\beta}$ by straightforward linear
response theory (see Supplemental Material \cite{supplemental}).

The excited spin density is a source term for the kinetic equations, from
which we can calculate spin transport, and when the spin-orbit
coupling is sufficiently weak, the spin current. For the spin dynamics in the
2DEG we need not only $\mathbf{s}_{l}(\boldsymbol{\rho},t)$ from
Eq.~(\ref{linear_response}), but also its time derivative (\textquotedblleft$I$\textquotedblright\ for interaction
representation)
\cite{Maekawa,spin_pumping1,linear_response_Ale,spintronics,Yu_InAs}, which is
derived in the Supplemental Material \cite{supplemental}:
\begin{align}
\left\langle \frac{\partial\mathbf{\hat{s}}_{I}\left(  \boldsymbol{\rho
},t\right)  }{\partial t}\right\rangle \equiv{\pmb {\mathcal{R}}}(t)
=\frac{\partial\mathbf{s}_{l}(\boldsymbol{\rho} ,t)}{\partial t}+\mu_{0}%
\gamma_{e}\mathbf{s}_{l}(\boldsymbol{\rho},t)\times\mathbf{H}(\boldsymbol{\rho
},t). \label{local_conservation}%
\end{align}
This relation recovers that in the conventional spin pumping
\cite{spin_pumping1,spin_pumping2} when replacing microwave field
$\mathbf{H}$ by the magnetization $\mathbf{m}$ at the interface in
Eqs.~(\ref{linear_response}) and (\ref{local_conservation}) \cite{Maekawa}, in
which case ${\pmb{\mathcal{R}}}(t)$ can be interpreted as the spin injection
rate across the interface or spin-current gradient.

When the spin-orbit coupling is negligible, the spin current operator is
defined through the commutator $-\nabla_{\boldsymbol{\rho}}\cdot
\hat{{\pmb {\cal J}}}_{I}=(i/\hbar)\left[  \hat{H}_{0}^{I},\mathbf{\hat{s}%
}_{I}(\boldsymbol{\rho},t)\right]  $, leading to
\cite{Maekawa,linear_response_Ale}
\begin{equation}
-\nabla_{\boldsymbol{\rho}}\cdot{\pmb {\cal J}}(\boldsymbol{\rho}%
)={\partial\mathbf{s}_{l}(\boldsymbol{\rho},t)}/{\partial t}+\mu_{0}\gamma
_{e}\mathbf{s}_{l}(\boldsymbol{\rho},t)\times\mathbf{H}(\boldsymbol{\rho},t),
\label{spin_conservation}%
\end{equation}
where ${\pmb{\mathcal{J}}}(\boldsymbol{\rho})$ is the spin current tensor with
elements $\mathcal{J}_{\alpha}^{\delta}(\boldsymbol{\rho},z)$ ($\alpha$ and
$\delta$ are the spatial and spin indexes).
Substituting Eq.~(\ref{linear_response}) into Eq.~(\ref{spin_conservation})
leads to DC and AC spin currents. When the susceptibility is well behaved at
low frequencies and assuming that the Fermi energy $E_{F}\gg\hbar\omega,$ we
may use the adiabatic approximation for the excited spin density
\cite{linear_response_pumping,linear_response_Ale,Mahan,linear_response_pumping,linear_response_pumping2}%
,%
\begin{align}
&s_{l,\alpha}(\boldsymbol{\rho},t)    =-\mu_{0}\gamma_{e}\sum_{\mathbf{k}%
}e^{i\mathbf{k}\cdot\boldsymbol{\rho}}\mathrm{\mathrm{\operatorname{Re}}}%
\chi_{\alpha\beta}(\mathbf{k},\omega\rightarrow0)H_{\beta}(\mathbf{k}%
,t)\nonumber\\
&  +\mu_{0}\gamma_{e}\sum_{\mathbf{k}}e^{i\mathbf{k}\cdot\boldsymbol{\rho}%
}\left.  \frac{\partial\mathrm{\operatorname{Im}}\chi_{\alpha\beta}%
(\mathbf{k},\omega)}{\partial\omega}\right\vert _{\omega\rightarrow0}%
\frac{d{H}_{\beta}(\mathbf{k},t)}{d t}. \label{adiabatic}%
\end{align}
In the long wavelength limit in which $\mathrm{\mathrm{\operatorname{Re}}}%
\chi_{\alpha\beta}(\mathbf{k}\rightarrow0,\omega\rightarrow0)$ is constant the
first, reactive term on the r.h.s. of Eq.~(\ref{adiabatic}) causes a pure AC
contribution $\sim\mathbf{\dot{H}}$ to the spin current through the first term
$\dot{\mathbf{s}}$ in the r.h.s. of Eq.~(\ref{spin_conservation}). The second,
dissipative term contributes to a DC spin current. We disregard the nearly
homogeneous external microwaves that excites the wire FMR since it does not contribute to the DC response and 
assume that a small static magnetic field that align the wire magnetization
 has a negligible effect on the 2DEG spins. The spin current injected
under the nanowire can be used as a boundary condition for a spin transport
theory \cite{Yu_spin_diffusion}.

The DC spin current is evaluated below for two model systems with large
$g$-factors, viz. the 2DEGs in narrow-gap semiconductor heterostructures and
topological surface states and for the 1DEG in the Supplemental
Material \cite{supplemental}.

\textit{Dipolar spin pumping}.---For the free electron gas, the spin
susceptibility is isotropic \cite{Mahan,Vignale},
\begin{equation}
\chi(\mathbf{k},\omega)=\frac{\hbar^{2}}{2}\sum_{\mathbf{q}}\frac
{f(\xi_{\mathbf{q}})-f(\xi_{\mathbf{k}+\mathbf{q}})}{\hbar\omega+i0_{+}%
+\xi_{\mathbf{q}}-\xi_{\mathbf{k}+\mathbf{q}}}, \label{susceptibility_free}%
\end{equation}
where $\xi_{\mathbf{k}}=\hbar^{2}k^{2}/(2m^{\ast})-\mu$ is the electron energy
with effective mass $m^{\ast},$ relative to the chemical potential $\mu$, and
$f(\xi_{\mathbf{k}})=\{\exp[\xi_{\mathbf{k}}/(k_{B}T)]+1\}^{-1}$ is the
Fermi-Dirac distribution at temperature $T$. In the microwave regime for the
nanowire, $\left\vert k_{x}\right\vert <2k_{F},$ where $k_{F}$ is the 2DEG
Fermi vector [in semiconductors $k_{F}=O\left(  \mathrm{nm}^{-1}\right)  $]
and $\mathrm{Re}\chi(|k_{x}|,\omega\rightarrow0)=m^{\ast}/\left(  \pi\hbar
^{2}\right) $ \cite{Vignale} and the reactive first term vanishes. The DC spin
current then reduces to
\begin{align}
{\pmb {\cal J}}_{x}^{\mathrm{DC}}(x)  &  =(\mu_{0}\gamma_{e})^{2}\int_{0}%
^{x}dx\sum_{k_{x}}e^{ik_{x}x}\left.  \partial_{\omega}%
\mathrm{\operatorname{Im}}\chi(|k_{x}|,\omega)\right\vert _{\omega
=0}\nonumber\\
&  \times\left\langle \dot{\mathbf{H}}(k_{x},t)\times\mathbf{H}%
(x,t)\right\rangle _{\mathrm{DC}},\label{pumping_general}%
\end{align}
where we used the symmetry relations ${\pmb {\cal J}}_{x}^{\mathrm{D}C}%
(k_{x})=-{\pmb {\cal J}}_{x}^{\mathrm{DC}}(-k_{x})$ and ${\pmb {\cal J}}%
_{x}^{\mathrm{DC}}(x=0)=0$ \cite{supplemental}. Assuming for the moment that
$\chi(|k_{x}|,\omega)\approx\chi(k_{\mathrm{ave}},\omega)$ with
$k_{\mathrm{ave}}\sim\pi/(2w)$, we obtain the simplified expression at FMR
\begin{align}
{\pmb {\cal J}_{x}^{\mathrm{DC}}(x)}  &  \approx\left.  -2(\mu_{0}\gamma
_{e})^{2}\omega_{\mathrm{K}}\partial_{\omega}\mathrm{\operatorname{Im}}%
\chi\left(  |k_{x}|\rightarrow k_{\mathrm{ave}},\omega\right)  \right\vert
_{\omega=0}\nonumber\\
&  \times\int_{0}^{x}dx^{\prime}\mathrm{\operatorname{Im}}\left[
\tilde{\mathbf{H}}^{\ast}(x^{\prime})\times\tilde{\mathbf{H}}(x^{\prime
})\right]  . \label{pumping_simple}%
\end{align}
Hence, the DC spin-current below the nanowire is (approximately) proportional
to the transverse spin of the magnetic field, implying transfer of the photon
spin angular momentum to the electron spin with an efficiency governed by
$\left.  \partial_{\omega}\mathrm{\operatorname{Im}}\chi\left(  |k_{x}%
|\rightarrow k_{\mathrm{ave}},\omega\right)  \right\vert _{\omega=0}$. The
spin current is polarized in the $-y$-direction, i.e., opposite to the
magnetization direction of the nanowire.

Since ${\pmb {\mathcal{J}}}_{x}(x)\propto\mathrm{sgn}(x)$, the excited spin is
not chiral, but flows into both directions on both sides of the nanowire as
indicated by the blue arrow in Fig.~\ref{model}, just as in conventional spin
pumping
\cite{spin_pumping1,linear_response_Ale,linear_response_pumping,linear_response_pumping2}%
. Although excited by the same field, this result is in stark contrast to the
magnon spin current \cite{Yu1,Yu2,nanowire,Springer_book} or the chiral energy
currents of surface plasmon polaritons excited by a rotating electric dipole
\cite{near_field,Petersen,nano_optics}, which are both unidirectional and flow
in half space. We can trace the different physics to the collective nature of
magnons/plasmons with a well-defined dispersion relation that in the present
geometry are symmetric in $k$-space, but of which a chiral dipolar field
selects only one. The susceptibility of the non-interacting electron gas, on
the other hand, is made up by a broad spectrum of electron-hole pair
excitations at the Fermi energy, and chirality vanishes in the integral over
wave vectors at a given frequency. Interestingly, chirality emerges for
metallic carbon nanotubes or interacting electrons in a quantum wire that
cross the magnetic wire at right angles \cite{supplemental}, because the spin
susceptibility in the 1DEG with a linear dispersion at the Fermi
energy such as metallic carbon nanotubes with (Tomonaga-Luttinger liquid) or
without interactions is singular \cite{Mahan,supplemental}
\begin{equation}
\chi(k,\omega)=\frac{\hbar kL}{2\pi}\left(  \frac{1}{\omega+i0_{+}-v_{F}%
k}+\frac{1}{\omega+i0_{+}+v_{F}k}\right)  ,
\end{equation}
where $L$ is the system length and $v_{F}$ is the Fermi velocity. By contour
integration and the chiral dipolar field with $\mathbf{H}(k_{x}>0,\omega
_{\mathrm{K}})=0$ for the right-circularly polarized spin waves,
\begin{equation}
\mathbf{s}_{l}(x,t)=\mathrm{Im}\left(-\frac{2\mu_{0}\gamma_{e}\omega_{\mathrm{K}}}{v_{F}^{2}}\mathbf{H}\left(
k_{-},\omega_{\mathrm{K}}\right)  e^{ik_{-}x-i\omega_{\mathrm{K}}t}\right),
\end{equation}
when $x<0$ but vanishes when $x>0$,
with $k_{-}=-{\omega_{\mathrm{K}}}/v_{F}$, implying that the excited spin
density lives only in half of the nanowire. The DC spin current vanishes when
$x>0$, but flows in the same half space $x<0$ with
\begin{equation}
{\pmb {\mathcal {J}}}_{x}(x,t)=-\mu_{0}\gamma_{e}\int_{0}^{x}dx^{\prime
}\left.  \mathbf{s}_{l}(x^{\prime},t)\times\mathbf{H}(x^{\prime},t)\right\vert
_{\mathrm{DC}},
\end{equation}
recovering the chiral excitation of a spin-density current \cite{supplemental}
found earlier in magnetic films. This example proves that quite generally
chiral excitation by dipolar radiation is not caused by a hidden symmetry, but
requires poles in the spin-susceptibility generated by degenerate\textit{
}electron-hole pairs, or the plasmon, magnon, phonon excitations of a rigid
ground state.

We now estimate the magnitude of the DC spin current and/or spin-injection
rate by the dipolar field from an excited magnetic nanowire. We choose a
symmetric QW with $s=20$~nm of a semiconductor with small effective mass such
as InSb with $m^{\ast}=0.015m_{e}$ \cite{g_factor_InAs} and electron density
$n_{e}=3\times10^{11}~\mathrm{cm}^{-2}$ (corresponding to a Fermi energy
$E_{F}\sim50$~meV and Fermi temperature 560~K), such that only the lowest band
is populated even at room temperature. The Dresselhaus-type spin-orbit
coupling with coefficient $\gamma_{D}=220$~$\mathrm{eV\mathring{A}}^{3}$
\cite{g_factor_InAs,Dresselhaus_InSb} causes a small correction $\gamma
_{D}(\pi/s)^{2} k_{F}\sim0.7$~$\mathrm{meV}\ll E_{F}$ that we disregard. The
$g$-factor of electron is $g_{e}=-36$ \cite{g_factor_InAs,g_factor_InSb}, but
the sign is not important here. At temperature $T=100$~K the system is
degenerate with subband splitting $\hbar^{2}(\pi/s)^{2}/(2m^{\ast}%
)=63$~$\mathrm{meV}\gg k_{B}T$. For a Co or CoFeB nanowire with $w=d=60$~nm
and $\mu_{0}M_{s}=1.2$~T \cite{Yu2}, we assume a coherent magnon density
$\rho_{m}=|\langle\hat{\alpha}_{k_{y}=0}\rangle|^{2}=10^{9}$ in
Eq.~(\ref{magnon_operator}) that corresponds to a transverse magnetization
amplitude $M_{x,y}\sim2\sqrt{2\gamma\hbar M_{s}\rho_{m}}m_{x,y}^{k_{y}=0}$,
i.e. a small precession cone angle $\sim3.2\times10^{-3}$ degrees that is
easily excited by FMR.

\begin{figure}[th]
	\hspace{-0.2cm}{\includegraphics[width=4.45cm]{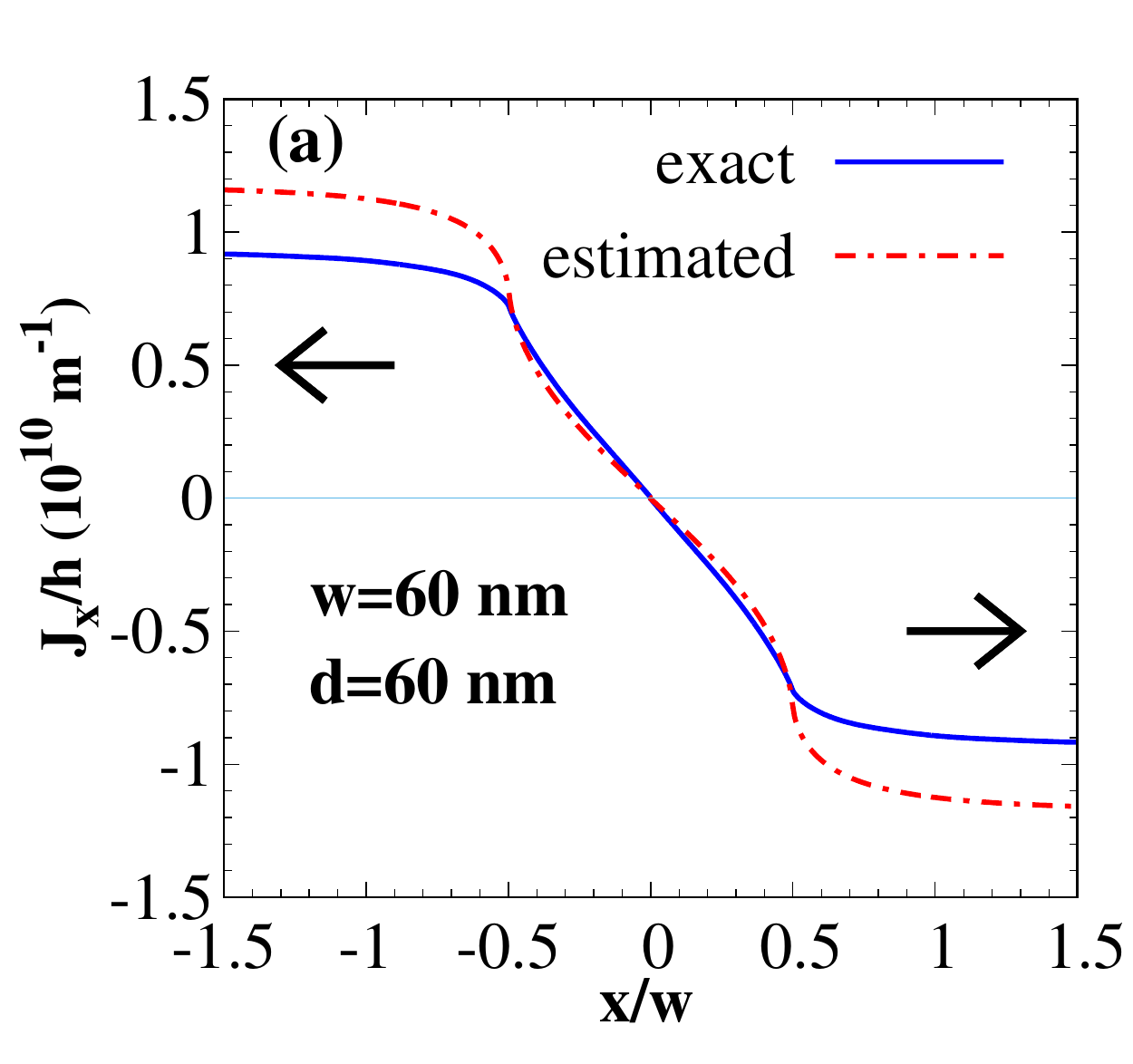}}
	\hspace{-0.2cm}{\includegraphics[width=4.45cm]{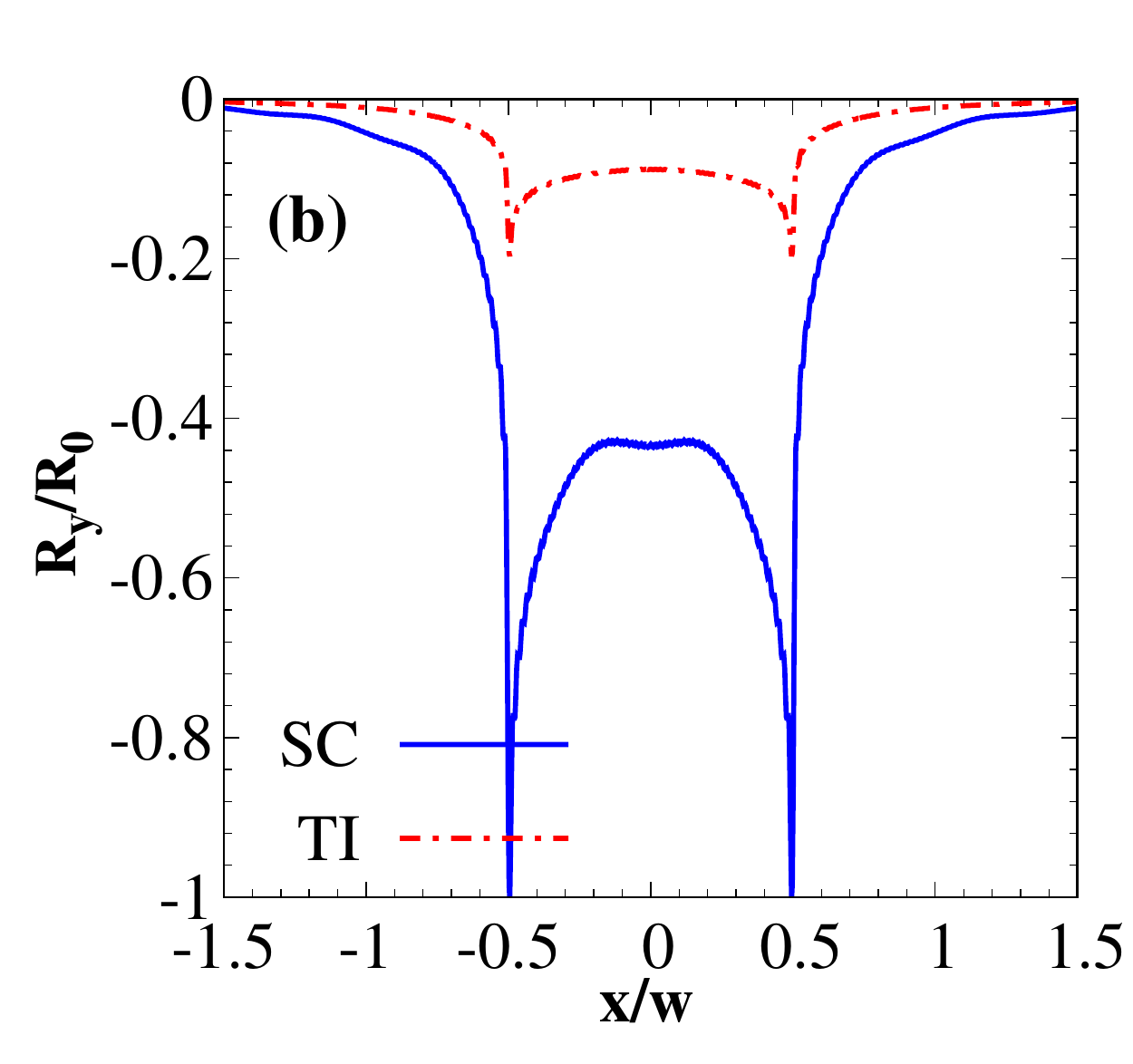}}
	\caption{Excited spin current ${\pmb {\mathcal{J}}}_{x}/h$,
		with $h=2\pi\hbar$ [(a)] and spin-injection rate [(b)] under an excited Co
		nanowire. In (a), the blue and red curves are calculated from
		Eqs.~(\ref{pumping_general}) and (\ref{pumping_simple}), respectively. In (b),
		the blue and red curves are the spin injection rates for a semiconductor 2DEG
		and the surface state of a topological insulator, normalized by the maximal
		magnitude $R_{0}/\hbar=2.7\times10^{18}$~m$^{-2}$ of the blue curve. }%
	\label{spin_current}%
\end{figure}

We plot the DC spin current under the nanowire from
Eqs.~(\ref{pumping_general}) and (\ref{pumping_simple}) in
Fig.~\ref{spin_current}(a), in which the bidirectional spin current is
indicated by the black arrows, and the spin-injection rate in
Fig.~\ref{spin_current}(b). The simplified Eq.~(\ref{pumping_simple}) describes the pumped
current by Eq.~(\ref{pumping_general}) well. With the same conditions, the
pumped spin current is four times in magnitude smaller in InAs QWs ($m^{\ast
}=0.023m_{e}$ and $|g_{e}|=14.3$ \cite{g_factor_InAs}). The spin current is of
the same order as the spin Hall current generated by an electric field of
$0.1~\mathrm{kV}/\mathrm{cm}$ and a spin Hall conductivity $\sigma_{x}%
^{y}=10^{6}~(\Omega\cdot\mathrm{m})^{-1},$ which should be easily measurable
\cite{SHE_RMP}.\textit{ }Under the same conditions, the spin current pumped by
the dipolar interaction is comparable with that from interfacial exchange
interaction with an exchange splitting $JM_{s}\hbar\sim10$~meV
\cite{estimation}, but does not require good electric contact between magnet
and semiconductor.

The excited spin current under the transducer drives diffusive spin transport
over the spin diffusion length scale \cite{Yu_spin_diffusion}. The spin signal
can be converted to a transverse voltage by the inverse spin Hall effect in
the 2DEG itself or by heavy metal contacts \cite{SHE_RMP} or the inverse
Edelstein effect \cite{IEE}. The cyclotron resonance excited by evanescent
microwave magnetic fields in the quantum Hall regime could be an interesting
extension of the present work.

Finally, we estimate the efficiency of the dipolar spin pumping for surface
states of the $n$-doped topological insulator Bi$_{2}$Se$_{3}$ \cite{Bi2Se3}
at a low temperature $T=30$~K for which a good exchange interaction with
magnetic contacts is difficult to achieve \cite{YIG_topology} and perhaps not
desired because of an associated proximity effect. Since the spin current is
not conserved, we focus on the DC spin injection rate ${\pmb {\mathcal{R}}}$
defined in Eq.~(\ref{local_conservation}) and compared with the semiconductor
case in Fig.~\ref{spin_current}(b). Only the diagonal terms of the
susceptibility tensor \cite{ss_3D}
\begin{equation}
\chi(|k_{x}|,\omega)=\frac{\hbar^{2}}{8}\sum_{\mathbf{q}}(1-\cos
\phi_{\mathbf{q}})\frac{n_{F}(\xi_{\mathbf{q}-\mathbf{k}})-n_{F}%
(\xi_{\mathbf{q}})}{\hbar\omega+i0_{+}+\xi_{\mathbf{q}-\mathbf{k}}%
-\xi_{\mathbf{q}}},
\end{equation}
where $\xi_{\mathbf{k}}=\hbar v_{F}k-\mu$, contribute to the DC spin injection
(see Supplemental Material \cite{supplemental}). With $n_{e}=10^{11}$%
~cm$^{-2}$, $v_{F}=10^{5}$~m/s, and $\left\vert g_{e}\right\vert =20$
\cite{g_factor_Bi2Se3}, the spin injection rate is of the same order as that
of InSb semiconductor 2DEG.

\textit{Discussion}.---Spin pumping by evanescent microwaves is a coherent
mechanism for the generation of pure spin currents in conventional spintronic
systems and devices that requires nanomagnets rather than extended films. The
effect is not small: spin currents generated by the stray fields of Co
nanowires on top of a thin yttrium iron garnet film exceed those by the
exchange coupling \cite{Yu2}. Here we focus on transverse spin
pumping into a conductor, which is most efficient for low-dimensional electron
systems. We predict here a transverse spin current density $10^{-13}%
~\mathrm{J/m}$ pumped into an InAs 2DEG, which is almost two orders of
magnitude larger than what has been observed for the spin-pumping by a Py slab
into graphene \cite{graphene_exp2}, whose signal should indeed be
much smaller with small $g$-factor. We therefore cannot exclude that
the observations are caused by dipolar fields at the edge of Py, and not
exchange interactions at interface. In general, however, spin
pumping into 2DEGs by extended magnetic films
\cite{GaAs_bulk,STO_LAO,topological_exchange,topological_exchange2} (and in
basically all planar structures used in conventional spin pumping experiments)
should be dominated by the exchange mechanism. 

The photon angular momentum is inherent to the evanescent stray fields of a
precessing magnetization, but it also exists in microwave cavities or
waveguides. The dipolar spin pumping is contactless and avoids possible
artifacts by the magnetic proximity effect. The excited spin current is not
chiral for 2DEGs, but chirality re-emerges  in the 1DEG. The spin pumping by a magnetic transducer into a 2DEG and a
surface state of a topological insulator are estimated large enough to be
observable. Our study bridges the concepts and understandings in different
fields including spintronics
\cite{spintronics,magnonics1,magnonics2,spintronics_RMP,spin_pumping2},
nano-optics \cite{nano_optics} and plasmonics \cite{Nori_Science,Nori_review}.

\vskip0.25cm \begin{acknowledgments}
This work is financially supported by the Nederlandse Organisatie voor Wetenschappelijk Onderzoek (NWO) as well as JSPS KAKENHI Grant No. 19H006450. We thank Yaroslav Tserkovnyak, Alejandro O. Leon and Jin Lan for useful discussions.
\end{acknowledgments}


\begin{thebibliography}{67}
\bibitem {footnote}For the Kittel mode, $m_{x}=i\sqrt{\mathcal{D}/(4wd)}$ and
$m_{z}=\sqrt{1/(4\mathcal{D} w d)}$, where $\mathcal{D}=\sqrt{\frac
{H_{0}+N_{zz}M_{s}}{H_{0}+N_{xx}M_{s}}}$ is determined by the applied field
$H_{0}$ and $M_{s}$ with $N_{xx}\simeq d/(d+w)$ and $N_{zz}\simeq w/(d+w)$. It
is circularly polarized $m_{z}=im_{x}$ at large magnetic fields ($H_{0}\gg
M_{s}$) and/or when the wire has a square cross section. It tends to be
linearly polarized at weak applied fields and when $w\gg d$.
\bibitem {estimation}The linearized interface exchange Hamiltonian $\hat
{H}_{\mathrm{ex}}=J\int\hat{\mathbf{s}}(\boldsymbol{\rho},t)\cdot
\hat{\mathbf{M}}(\mathbf{r},t)d\boldsymbol{\rho}$, with $\hat{\mathbf{M}%
}\simeq\left(  m_{x}(x),M_{s},m_{z}(x)\right)  ^{T}$. With $\mathbf{m}%
(x,t)=\tilde{\mathbf{m}}(x,\omega_{\mathrm{K}})e^{-i\omega_{\mathrm{K}}%
t}+\tilde{\mathbf{m}}^{\ast}(x,\omega_{\mathrm{K}})e^{i\omega_{\mathrm{K}}t}$,
we arive at the estimate
\begin{align}
{\pmb {\cal J}_{x}^{\mathrm{DC}}(x)}  &  \approx-2J^{2}\omega_{\mathrm{K}%
}\partial_{\omega}\mathrm{\operatorname{Im}}\chi\left(  \left\vert
k_{x}\right\vert \rightarrow\pi/(2w),\omega\right)  |_{\omega=0}\nonumber\\
&  \times\int_{0}^{x}dx^{\prime}\mathrm{\operatorname{Im}}\left[
\tilde{\mathbf{m}}^{\ast}(x^{\prime})\times\tilde{\mathbf{m}}(x^{\prime
})\right]  . \nonumber\label{pumping_exchange}%
\end{align}
\end{thebibliography}
\end{document}